\documentclass[prl,twocolumn,showpacs,floatfix,citeautoscript,
               superscriptaddress]{revtex4}
\usepackage{graphicx}
\usepackage{bm}
\usepackage{amsmath}

\begin{document}
\title{A quantum criticality perspective on the
       charging of narrow quantum-dot levels}

\author{V. Kashcheyevs}
\affiliation{Faculty of Physics and Mathematics,
                  University of Latvia, Ze{\c{l}\c{l}}u
                  street 8, Riga LV-1002, Latvia}
\author{C. Karrasch}
\affiliation{Institut f{\"u}r Theoretische Physik A,
                  RWTH Aachen University and
                  JARA---Fundamentals of Future Information
                  Technology, 52056 Aachen, Germany}
\author{T. Hecht}
\author{A. Weichselbaum}
\affiliation{Department of Physics, Arnold Sommerfeld
                  Center for Theoretical Physics and
                  Center for NanoScience,
                  Ludwig-Maximilians-Universit\"at,
                  80333 Munich, Germany}
\author{V. Meden}
\affiliation{Institut f{\"u}r Theoretische Physik A,
                  RWTH Aachen University and
                  JARA---Fundamentals of Future Information
                  Technology, 52056 Aachen, Germany}
\author{A. Schiller}
\affiliation{Racah Institute of Physics, The Hebrew
                  University, Jerusalem 91904, Israel}

\begin{abstract}
Understanding the charging of exceptionally narrow
levels in quantum dots
in the presence of interactions remains a
challenge within mesoscopic physics. We address this
fundamental question in the generic model of a narrow
level capacitively coupled to a broad one. Using
bosonization we show that for arbitrary capacitive
coupling charging can be described by an analogy to
the magnetization in the anisotropic Kondo model,
featuring a low-energy crossover scale that depends
in a power-law fashion on the tunneling amplitude
to the level. Explicit analytical expressions for
the exponent are derived and confirmed by detailed
numerical and functional renormalization-group
calculations.
\end{abstract}

\maketitle

\textit{Introduction.}
Confined nanostructures offer a unique arena for
thoroughly interrogating the interplay between
interference and interactions while holding the
promise of future applications. Particularly appealing
are semiconductor quantum dots (QDs), for which the
manipulation of spin~\cite{Craig-etal-04,Petta-etal-05}
and charge~\cite{PJMHG04} has recently been demonstrated.
The precise and rapid control of switchable gate
voltages renders these devices attractive candidates
for a solid-state qubit~\cite{LDiV98,EL05}. The accurate
manipulation of QD setups requires, however, detailed
understanding of how charging proceeds. Indeed,
interactions can substantially modify the orthodox
picture of charging, whether by renormalizing the
tunneling rates or by introducing nonmonotonicities
into the population of individual
levels~\cite{SI00,KG05,SSOvD05}. Even the simplest
two-level device, where each level harbors only a
single spinless electron, displays remarkably rich
behavior~\cite{Special-NJP-issue}.

We consider a situation in which the width of one
narrow level is much smaller than the width of
the other broad one. A disparity in widths
is generic for QDs in the intermediate regime
between integrable and chaotic~\cite{SI00}.
It was reported in several artificial
structures~\cite{Ensslin02,Kobayashi04}, and has been
exploited for charge sensing~\cite{JMHG04,BvOG05}.
As the energy $\epsilon_{-}$ of the narrow level is
raised,
its occupation varies from $1$ to $0$
over a characteristic width $\Omega$.
This energy scale, or the corresponding
charge-fluctuation time scale $\hbar/\Omega$,
manifests itself in charge sensing and transmission-phase
measurements~\cite{Phase-lapses}. The effect of
inter-level repulsion $U$ on $\Omega$ has been
explored only in the large-$U$ limit, revealing novel
correlation effects~\cite{MM06,KSAE07,LK07,SI07}.
The physical mechanism determining $\Omega$
for moderate $U$
remains unclear~\cite{Special-NJP-issue}.

In this Letter we solve the fundamental question of the
charging of a narrow QD level from a quantum-critical
perspective. Due to the capacitative coupling $U$,
every switching of the narrow level
initiates restructuring of the broad level and its attached Fermi sea,
in direct analogy
with the x-ray edge singularity.
For nonzero tunneling to the narrow level,
coherent superpositions of these charge re-arrangements
lead to Kondo physics~\cite{AYH70} with the charge state
($0$ or $1$)
acting as
a pseudo-spin, and the
energy of the narrow level acting as a
Zeeman field.
Using Abelian bosonization we show that $\Omega$, being
the Kondo scale in the pseudo-spin language, depends
on the tunneling amplitudes in a power-law fashion. We
derive explicit analytical expressions for the exponents
encompassing all physical regimes of the model (at zero
temperature $T$). In a second step we confirm our
predictions by detailed numerical~\cite{NRG} (NRG)
and functional~\cite{FRG} (FRG) renormalization-group (RG)
calculations, thus resolving this challenging aspect of
mesoscopic physics.

\textit{Model and objective}.
Our specific model for charging is depicted schematically
in the inset of Fig.~1, and is defined by the
Hamiltonian ($\sigma$ is the pseudo-spin index)
\begin{align}
 {\cal H} & =   \sum_{\sigma = \pm}  \Bigl [
             \sum_k
                \epsilon^{}_k
                c^{\dagger}_{k \sigma} c^{}_{k \sigma}
          + V_{\sigma} \sum_k
                \bigl (
                         c^{\dagger}_{k \sigma} d^{}_{\sigma}
                         + d^{\dagger}_{\sigma} c^{}_{k \sigma}
                \bigr )
\nonumber \\
       &
           + \epsilon^{}_{\sigma}
                          d^{\dagger}_{\sigma} d^{}_{\sigma} \bigr ]     
          + b/2 \bigl (
                          d^{\dagger}_{+} d^{}_{-}
                          + d^{\dagger}_{-} d^{}_{+}
                        \bigr )
          + U \Delta\hat{n}_+ \Delta\hat{n}_-  .
\label{H-A}
\end{align}
Here, $d^{\dagger}_{\pm}$ ($c^{\dagger}_{k \pm}$)
creates an electron on the dot (in the leads),
and $\Delta\hat{n}_\pm$ equals
$d^{\dagger}_{\pm} d^{}_{\pm} - 1/2$. Equation
\eqref{H-A} is a generalized Anderson impurity model
with pseudo-spin-dependent tunneling amplitudes
$V^{}_{+} \geq V^{}_{-} \geq 0$ and a tilted
magnetic field,
whose components are
$ \epsilon^{}_+ - \epsilon^{}_- $
and the direct hopping amplitude $b$.
This form
follows from a generic model of spinless
electrons with two
dot
levels and two leads by
simultaneous
unitary transformations in the dot and the
lead space~\cite{LK07,KSAE07,SI07}.
The Hamiltonian~\eqref{H-A}
has recently gained considerable attention in
connection with phase lapses, population inversion,
and many-body resonances~\cite{Special-NJP-issue}.
The energies $\epsilon^{}_{\pm}$ are tuned using
gate voltages. Depending on the specific realization,
their tuning
may inflict a similar
change in $b$. We 
focus on realizations where
$\epsilon^{}_{\pm}$ can be tuned independently of $b$.
\begin{figure}
\includegraphics[height=4.60cm,clip]{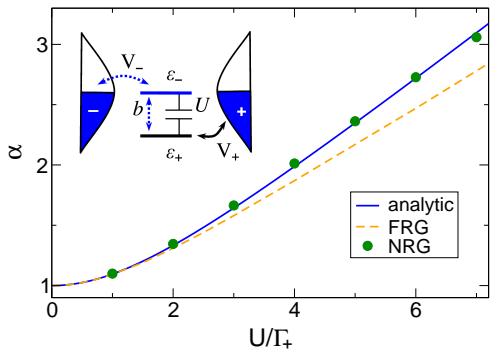}
\caption{(Color online) The exponent $\alpha$ computed
         using the NRG, FRG, and Eq.~(\ref{powers}).
         NRG parameters: $\Gamma_{+}/D = 0.04$,
         $\Lambda = 1.7$, and 2800 states are retained.
         Inset: The model system. Two localized QD
         levels are coupled by tunneling to separate
         baths. Spinless electrons residing on the
         two levels experience a Coulomb repulsion $U$.
} \label{fig:fig1}
\end{figure}

The bare energy scales that
characterize
tunneling in Eq.~(\ref{H-A}) are the level broadenings
$\Gamma_{\pm} = \pi \rho V_{\pm}^2$ and the direct
hopping amplitude $b$. The density of states (DOS)
$\rho$ is taken
to be
equal for both bands without loss
of generality. Our interest is in the charging properties
of the narrow level $d^{\dagger}_{-}$ as a function of
$\epsilon^{}_{-}$ in the limit where $b$ and $V^{}_{-}$
are both small: $\Gamma_{-}, b \ll \Gamma_{+}$.
Strictly at $b \!= \!V^{}_{-}\!\! =\! 0$
ergodicity of the microcanonical
ensemble is broken as a new conserved quantity
arises: $\hat{n}^{}_{-} \!\!\equiv\!
d^{\dagger}_{-} d^{}_{-}$ is either equal to $0$ or $1$.
Comparing the total energies of the competing
ground states
with $\langle \hat{n}^{}_{-} \rangle = 0$ and
$\langle \hat{n}^{}_{-} \rangle = 1$ as a function
of $\epsilon^{}_{-}$ one finds a critical value
$\epsilon^{}_{-}=\epsilon^{\ast}(\epsilon_{+}, U, V_{+})$
at which the two become degenerate.
For $\epsilon^{}_{+} = 0$, $\epsilon^{\ast}$ is
pinned to zero by particle-hole symmetry if
symmetric bands are assumed.
In the limit $b, V_{-} \to 0$ the average occupation
$\langle \hat{n}^{}_{-} \rangle$
thus indicates a first-order transition (width $\Omega = 0$)
as $\epsilon^{}_{-}$ is swept across $\epsilon^{\ast}$.
It is the smoothening ($\Omega > 0$) of this
transition at small but finite $b, V_{-}$
for $U \geq 0$
that is addressed in
this
Letter.

Two regimes can be
distinguished depending on $\epsilon_{+}$.
When $|\epsilon^{}_{+}| \gg U, \Gamma_{+}$,
the level $d^{\dagger}_{+}$ maintains an approximately
fixed integer valence $\langle n^{}_{+} \rangle
\in \{ 0, 1 \}$, independent of $\epsilon^{}_{-}$.
Hence, the charging of $d^{\dagger}_{-}$ is essentially
single-particle in nature with $\Omega = \Gamma_{-}
+ \Gamma_{+} b^2/\epsilon_{+}^{2}$.
The effect of interactions is contained in the simple
Hartree renormalization, $\epsilon^{}_{-} \to
\epsilon^{}_{-} + U ( \langle n^{}_{+} \rangle - 1/2)$.
Far more complex is the case of $|\epsilon^{}_{+}|
\ll \max \{U, \Gamma_{+}\}$, when the broad level
is prone to strong valence fluctuations (for
$\epsilon^{}_{-} \! \to \! \epsilon^{\ast}$). Going from
$U/\Gamma_{+} \ll 1$ to $1 \ll U/\Gamma_{+}$
spans all physical regimes from weak to strong
electronic correlations~\cite{KSAE07,LK07,SI07},
which constitutes
the main focus of our study.
To this end we initially
set $\epsilon^{}_{+} = 0$, which fixes
$\epsilon^{\ast} = 0$. Using analytical and numerical
tools we first obtain $\Omega$ in the case where either
$V^{}_{-}$ or $b$ is nonzero. The combined effect of
$V^{}_{-}$ and $b$ is
next
addressed by single-parameter scaling and FRG. Finally,
we extend our
analytical results
to arbitrary $\epsilon_{+}$.

\textit{Analytical approach}.
To analytically determine the width $\Omega$ using
minor approximations, we proceed in two steps. First,
we derive a continuum-limit Hamiltonian where
$\Gamma_{+}$ is incorporated in full. Second,
an exact mapping of this Hamiltonian onto the
anisotropic Kondo model is established. This allows
usage of known results for the Kondo problem in
order to extract $\Omega$.

In the first step, we diagonalize the Hamiltonian
${\cal H}_{+} = \sum_{k} \epsilon^{}_k
                        c^{\dagger}_{k +} c^{}_{k +}
             + V^{}_{+} \sum_k
               \bigl \{
                        c^{\dagger}_{k +} d^{}_{+}
                        + d^{\dagger}_{+} c^{}_{k +}
               \bigr \} $
using scattering theory.
Expanding $d^{\dagger}_{+}$ in terms of the
single-particle eigen-modes of ${\cal H}_{+}$ and
converting to continuous constant-energy-shell
operators~\cite{LSA03}, ${\cal H}$ takes the form
of a generalized interacting resonant-level model
with a single $d^{\dagger}_{-}$ level tunnel
coupled to two bands: a narrow $\sigma = +$
band with a Lorentzian DOS of half-width
$\Gamma_{+}$, and a flat $\sigma = -$ band
with half-width $D \gg \Gamma_{+}$. In addition,
the $d^{\dagger}_{-}$ level is capacitively coupled
to the `$+$' band.

In the desired limit $b, \Gamma^{}_{-} \ll \Gamma^{}_{+}$,
one can conveniently replace the Lorentzian DOS with a
flat symmetric one of height $1/\pi \Gamma_{+}$ and
half-width $D_{+} = \pi \Gamma_{+}/2$~\cite{LSA03}.
The elimination of all degrees of freedom in the
energy interval $D_{+} < |\epsilon| < D$ leads to
renormalizations of the couplings of the order of
$\Gamma^{}_{-}/\Gamma^{}_{+} \ll 1$ or higher,
which can
be safely  neglected. Converting at this point
to left-moving fields, we obtain the continuum-limit
Hamiltonian
\begin{multline}
{\cal H} = i \hbar v_F \sum\nolimits_{\sigma = \pm}
               \int_{-\infty}^{\infty}
                     \psi^{\dagger}_{\sigma}(x)
                     \partial_x \psi^{}_{\sigma}(x) dx
          + \epsilon^{}_{-} d^{\dagger}_{-} d^{}_{-}
\\
          + (b/2) \sqrt{a}
            \bigl \{
                     \psi^{\dagger}_{+}(0) d^{}_{-}\!
                     + \text{H.c.}
            \bigr \}
          + U a :\! \psi^{\dagger}_{+}(0) \psi_{+}(0)\!\!:
            \!\Delta\hat{n}_{-}
\\
          + \sqrt{ a \Gamma_{+}\Gamma_{-}}   
            \bigl \{
                     \psi^{\dagger}_{-}(0) d^{}_{-}\!
                     + \text{H.c.}
            \bigr \} \, ,
\label{H-CL}
\end{multline}
applicable at energies below $\Gamma_{+}$. Here,
$a = \pi \hbar v_F/D_{+}$ is a new short-distance
cutoff (``lattice spacing''), and
$:\! \psi^{\dagger}_{+} \psi_{+}\!\!:$ stands for
normal ordering with respect to the filled Fermi sea.
The left-moving fields obey canonical anticommutation
relations subject to the regularization
$\delta(0) = 1/a$. The derivation of Eq.~(\ref{H-CL})
is controlled by the small parameters
$\Gamma^{}_{-}/\Gamma^{}_{+} \ll 1$ and
$b/\Gamma_{+} \ll 1$, and hence is expected to become
asymptotically exact as $\Gamma_{-}, b \to 0$.

If either $b = 0$ or $\Gamma_{-} = 0$,
Eq.~(\ref{H-CL}) can be treated using Abelian
bosonization~\cite{Haldane81}. To this end, we introduce
two bosonic fields $\Phi_{\pm}(x)$, one for each fermion
field $\psi_{\pm}(x)$. With a proper choice of the
phase-factor operators, the bosonized Hamiltonian reads
\begin{multline}
\label{H-bosonized}
{\cal H} = \sum\nolimits_{\sigma = \pm}
             \frac{\hbar v_F}{4 \pi}
             \int_{-\infty}^{\infty}
                  [ \nabla \Phi_{\sigma}(x) ]^2 dx
          + \epsilon^{}_{-} d^{\dagger}_{-} d^{}_{-}
\\
          + \hbar v_F \frac{2 \delta_{U}}{\pi}
            \nabla \Phi_+(0) \Delta\hat{n}_{-}
          + \frac{A}{\sqrt{2}}
                  \Bigl \{
                           e^{i \Phi_{\pm}(0)} d^{}_{-}
                           + {\rm H.c.}
                  \Bigr \},
\end{multline}
The tunneling term in Eq.~(\ref{H-bosonized}),
proportional to $A$, depends on the case of
interest; one takes $A = b/2$ and the upper sign
($A = \sqrt{\Gamma_{+}\Gamma_{-}}$,
lower sign) for $\Gamma_{-} = 0$ ($b = 0$).
The value of $\delta_{U}=\arctan (U/2 \Gamma_{+})$ is fixed by matching the
$b = \Gamma_{-} = 0$
scattering phase shifts
of the `+' band in
the fermionic and the bosonic representations,
for
each sector with
fixed integer occupancy of the `$-$' level.

Next, we manipulate
Eq.~(\ref{H-bosonized}) by (i) applying the canonical
transformation ${\cal H}' = \hat U^{\dagger} {\cal H}
\hat U$ with
\begin{equation}
       \hat U = \exp
                \left [
                       - i (2 \delta_U/\pi) \Phi_{+}(0)
                       \Delta\hat{n}_{-}
                \right ] \, ,
\label{Canonical-trans}
\end{equation}
and (ii) converting to the ``spin'' and ``charge''
fields $\Phi_{s}(x)$ and $\Phi_{c}(x)$. The latter
are defined as $\Phi_{s}(x) = \Phi_{+}(x)$ and
$\Phi_{c}(x) = \Phi_{-}(x)$ for $\Gamma_{-} = 0$,
and
\begin{gather}
\Phi_{s, c}(x)\! =\!
                     \bigl [ {1 +
                            (
                                    2 \delta_U/\pi
                            )^2} \bigr ]^{-1/2}
                 \Bigl [
                         \Phi_{\mp}(x) \mp
                         \frac{2 \delta_U}{\pi} \Phi_{\pm}(x)
                 \Bigr ]
\end{gather}
for $b = 0$ (the upper signs correspond to $\Phi_s$).
In this manner, the Hamiltonian acquires the unified
form
\begin{eqnarray}
{\cal H}' &=& \sum\nolimits_{\mu = s, c}
              \frac{\hbar v_F}{4 \pi}
              \int_{-\infty}^{\infty}
                   [ \nabla \Phi_{\mu}(x) ]^2 dx
           + \epsilon^{}_{-} d^{\dagger}_{-} d^{}_{-}
\nonumber \\
        && + \frac{A}{\sqrt{2}}
                     \bigl \{
                      e^{i \gamma \Phi_{s}(0)} d^{}_{-}
                      + d^{\dagger}_{-}
                        e^{-i \gamma \Phi_{s}(0)}
             \bigr \} ,
\label{H'}
\end{eqnarray}
where $\gamma = \sqrt{1 + (2 \delta_U/\pi)^2}$
for $b = 0$ and $\gamma = 1 - 2 \delta_U/\pi$
for $\Gamma_{-} = 0$.

The very same Hamiltonian with
$0 < \gamma < \sqrt{2}$ also describes the anisotropic
Kondo model with $0 < J_z$, where in standard notation
$A = J_{\perp}/\sqrt{8}$ and
$\gamma = \sqrt{2}
         \bigl [
                 1 - (2/{\pi})
                     \arctan (
                                     \pi \rho J_z/4
                             )
         \bigr ]$
represent the transverse and longitudinal spin-exchange
couplings respectively, and $\epsilon^{}_{-} = \mu_B g H$ corresponds
to a local magnetic field. This representation of
the Kondo model is obtained by \cite{Schlottmann80II}
(i) bosonizing the
Kondo Hamiltonian with two bosonic fields
$\Phi_{\uparrow}(x)$ and $\Phi_{\downarrow}(x)$, (ii)
converting to the spin and charge fields $\Phi_{s, c}(x)
= [\Phi_{\uparrow}(x) \mp \Phi_{\downarrow}(x)]/\sqrt{2}$,
(iii) 
employing
${\cal H}' = \hat{T}^{\dagger} {\cal H} \hat{T}$
with $\hat{T} = \exp [ -i \sqrt{2} (2 \delta_z/\pi)
\Phi_{s}(0) \tau_z ]$, $\tau_z$ being the $z$ spin
component and $\delta_z = \arctan ( \pi \rho J_z/4)$,
and (iv) representing the spin $\vec{\tau}$ in terms
of the fermion $d^{}_{-} = \tau^{-}$. This establishes
a mapping between our problem with either $b = 0$
or $\Gamma_{-} = 0$ and the anisotropic Kondo model.
In particular, charging of the $d^{\dagger}_{-}$ level
is mapped onto the magnetization of the Kondo impurity,
relating the width $\Omega$ to the Kondo temperature
$T_K$.

We can now exploit known results for the Kondo problem.
Specifically,
RG equations perturbative in $J_{\perp}$
but nonperturbative in $J_{z}$~\cite{AYH70}
give $T_K \sim D_{+} (A/D_{+})^{2/(2 - \gamma^2)}$,
which yields for our problem
\begin{gather}
\frac{\Omega}{\Gamma_{+}}
       \sim
          \begin{cases}
              (\Gamma^{}_{-}/\Gamma^{}_{+})^{\alpha}
              & \text{if } b = 0 \, , \\
              (b/\Gamma_{+})^{2 \beta}
              & \text{if } \Gamma_{-} = 0 \, ,
          \end{cases}
\\
\alpha = \frac{1}
              {1 - (2\delta_U /\pi)^2} \; ,
\;\;\;\;
\beta = \frac{1}
             {2 - \left [
                          1 - (2\delta_U /\pi)
                  \right ]^2} \; .
\label{powers}
\end{gather}
Thus, $\Omega$ is a power law of the relevant tunneling
amplitude with an exponent that varies smoothly with
$U$. In going from $U = 0$ to $U \gg \Gamma_{+}$,
$\alpha$ grows monotonically from $1$ to
$\pi U/(8 \Gamma_{+})$ while
$\beta$ decreases from $1$ to $1/2$. The asymptote
$\alpha = \pi U/(8 \Gamma_{+})$
coincides with
the result of Ref.~\onlinecite{KSAE07} [Eq.~(29) with
$\epsilon_0 = -U/2$], obtained using very different
techniques.
For $U = 0$, the noninteracting integer exponents are
reproduced.
Hence Eqs.~(\ref{powers}) are precise
both at small and large $U$. As shown next, these
expressions remain highly accurate also at intermediate
$U$, suggesting that they might actually be exact.

\begin{figure}
\includegraphics[width=7cm,clip]{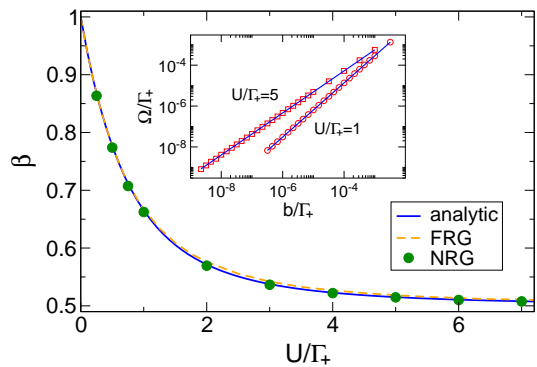}
\caption{(Color online) The exponent $\beta$ computed
         using the NRG, FRG, and Eq.~(\ref{powers}).
         NRG parameters: $\Gamma_{+}/D = 0.02$,
         $\Lambda = 1.6$, and 2000 states retained.
         Inset: representative NRG data for $\Omega$
         vs $b$, along with the log-log fits used to
         extract $\beta$.
} \label{fig:fig2}
\end{figure}

\textit{Numerical analysis}.
To test Eqs.~(\ref{powers}), we computed $\alpha$
and $\beta$ numerically using the NRG~\cite{NRG} and
FRG~\cite{FRG}, each approach having its own distinct
advantage. The NRG is extremely accurate in all parameter
regimes of interest, while the FRG is approximative
in $U$ but offers a far more flexible framework for
scanning parameters. The width $\Omega = 1/(\pi \chi_c)$
was obtained with either method from the inverse charge
susceptibility $\chi_c = d\langle \hat{n}^{}_{-}
\rangle/ d \epsilon^{}_{-}$, evaluated at
$\epsilon^{}_{-} = 0$ and  $T \to 0$. The
exponents $\alpha$ and $\beta$ were extracted from
log-log fits (see the inset to Fig.~\ref{fig:fig2}).
Our results, summarized in Figs.~\ref{fig:fig1} and
\ref{fig:fig2}, reveal excellent agreement between
Eqs.~(\ref{powers}) and the NRG, to within numerical
precision. The agreement extends to all interaction
strengths from small to large $U$, confirming the
accuracy of Eqs.~(\ref{powers}) at all $U$. The
FRG results for $\alpha$ coincide with those of
the NRG up to $U/\Gamma_{+} \approx 2$, above
which they acquire a linear slope that is reduced
by a factor of $8/\pi^2$ as compared to the
NRG~\cite{KSAE07}. The exponent $\beta$ is accurately
reproduced up to larger values of $U/\Gamma_{+}$.
In particular, the FRG data for $\alpha$ and $\beta$
exactly reproduce the leading behaviors of
Eqs.~(\ref{powers}) at small $U$.

\begin{figure}
\includegraphics[width=7cm,clip]{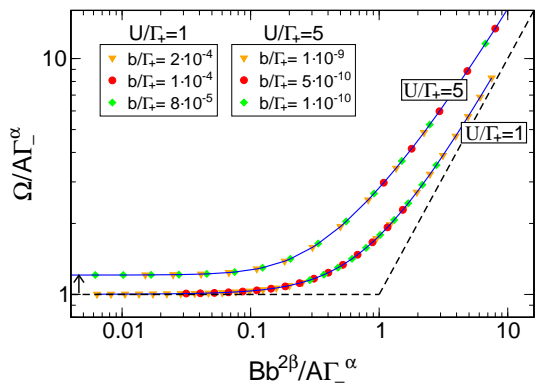}
\caption{(Color online) A scaling plot of $\Omega$
         for two (fixed) ratios
         $U/\Gamma_{+}$ and different combinations
         of $\Gamma_{-}$ and $b$, obtained using
         the FRG. The coefficients $A$ and $B$ were
         extracted from the limiting cases where
         $b = 0$ and $\Gamma_{-} = 0$,
         respectively~\cite{Comment-on-coefficients}.
         For clarity, the data for $U/\Gamma_{+} = 5$
         were multiplied by a constant as indicated by
         the arrow length. Dashed lines show the
         asymptotes ${\cal F} = 1$ and ${\cal F} = x$.}
\label{fig:fig3}
\end{figure}

\textit{Combination of $\Gamma_{-}$ and $b$}.
The case where both $\Gamma_{-}$ and $b$ are nonzero
lies beyond the scope of our bosonization treatment,
but allows the formulation of a scaling law.
To this end, consider the dimensionless quantity
$\tilde{\Omega} = \Omega/D_{+}$, which depends
on the three dimensionless parameters in Eq.~(\ref{H-CL}):
$\tilde{\Omega} = f(\tilde{V}, \tilde{b}, \delta_{U})$,
with $\tilde{V} = \sqrt{\Gamma_{+}\Gamma^{}_{-}}/D_{+}$
and $\tilde{b} = b/D_{+}$. Given the exact
RG
trajectories, $\tilde{\Omega}$
evolves according to $\tilde{\Omega}' = \tilde{\Omega}/\xi
= f(\tilde{V}', \tilde{b}', \delta'_{U}, \{\lambda'_i\})$
upon reducing the bandwidth from $D_{+}$ to $\xi D_{+}$
($0 < \xi < 1$). Here, primes denote renormalized
parameters and $\{\lambda'_i\}$ are the new couplings
generated. At sufficiently weak tunneling the RG equations
can be linearized with respect to the relevant couplings
$\tilde{V}'$ and $\tilde{b}'$, resulting in their
power-law growth with the exponents determined
previously: $\tilde{V}' = \tilde{V} \xi^{-1/2\alpha}$
and $\tilde{b}' = \tilde{b} \xi^{-1/2\beta}$. Note that
$\delta_{U}$ is left unchanged in this approximation,
nor are there any new couplings generated. Consequently,
$f(\tilde{V}, \tilde{b}, \delta_{U}) = \xi f(\tilde{V}
\xi^{-1/2\alpha}, \tilde{b} \xi^{-1/2\beta}, \delta_{U})
= \tilde{\Omega}$ is a homogeneous function of $\xi$,
taking the general form
$f(\tilde{V}, \tilde{b}, \delta_{U}) =
\tilde{V}^{2\alpha}
     {\cal G}
     (\tilde{b}^{2\beta}/\tilde{V}^{2\alpha},\delta_U)$.
Finally, defining the coefficients $A$ and $B$ from
$\Omega|_{b = 0} = A \Gamma^{\alpha}_{-}$ and
$\Omega|_{\Gamma_{-} = 0} = B b^{2 \beta}$, we arrive
at the scaling form~\cite{Comment-on-coefficients}
\begin{equation}
\Omega = A \Gamma^{\alpha}_{-}
         {\cal F}( B b^{2 \beta}/ A \Gamma^{\alpha}_{-};
                   \delta_U) ,
         \label{Scaling-form}
\end{equation}
with ${\cal F}( 0; \delta_U) = 1$ and
${\cal F}(x \gg 1; \delta_U) = x$. In
Fig.~\ref{fig:fig3} we confirm the scaling form
of Eq.~\eqref{Scaling-form} using FRG data.

\textit{Extension to arbitrary $\epsilon_{+}^{}$}.
Our discussion has focused thus far on
$\epsilon^{}_{+} = 0$. A nonzero $\epsilon^{}_{+}$
introduces the potential-scattering term
${\cal H}_{\rm ps} =
\epsilon^{}_{+} a \! :\! \psi^{\dagger}_{+}(0)
\psi_{+}(0)\!\!:$ into Eq.~(\ref{H-CL}). Consequently,
$\delta_{U}$ in Eq.~(\ref{H-bosonized}) is replaced
with two distinct parameters $\delta_{\pm} =
\arctan [(U \pm 2 \epsilon^{}_{+}) /2 \Gamma_{+}]$,
assigned to $\Delta\hat{n}_{-} = \pm 1/2$,
respectively. An identical derivation, only with
$2 \delta_U  \tilde{n}^{}_{-} \to
(\delta_{+} + \delta_{-}) \Delta\hat{n}_{-} +
(\delta_{+} - \delta_{-})/2$ in
Eq.~(\ref{Canonical-trans}), leads then to the
same Hamiltonian (\ref{H'}) with two modifications:
(i) $\epsilon^{}_{-}$, and thus $\epsilon^{\ast}$,
acquires a shift proportional to
$\delta^{2}_{+} - \delta^{2}_{-}$, and (ii)
$\delta_{U}$ is replaced with
$(\delta_{+} + \delta_{-})/2$ in the expressions
for $\gamma$. The end results for $\alpha$ and
$\beta$ are just Eqs.~(\ref{powers}) with
$\delta_{U} \to (\delta_{+} + \delta_{-})/2$,
which properly reduce to the noninteracting limit
$\alpha = \beta = 1$ when
$|\epsilon^{}_{+}| \gg U, \Gamma_{+}$. The effect
of nonzero $\epsilon^{}_{+}$ is negligible for
$|\epsilon^{}_{+}| \ll \max \{ U, \Gamma_{+} \}$.
It becomes significant only as $|\epsilon^{}_{+}|$
approaches $\max \{ U, \Gamma_{+} \}$.

\textit{Summary}.
We have resolved the fundamental question of the
charging of a narrow QD level capacitively coupled
to a broad one. The zero-tunneling fixed point is
critical in the sense of being unstable. Finite
tunneling is a relevant perturbation, driving the
system to a strong-coupling Fermi-liquid fixed point.
The inverse charge-fluctuation time $\Omega$ varies
as a power of the bare tunneling amplitude, with a
nonuniversal exponent that depends on the nature of
tunneling, the strength of the capacitive
coupling, and the width and position of the broad
level. We have proven this scenario by devising a
two-stage mapping of the original model
onto the anisotropic Kondo problem, yielding
accurate analytic expressions for the exponents.
Our analytic predictions were confirmed by extensive
numerical calculations within the frameworks of the
NRG and FRG.

\textit{Acknowledgments}.
We thank A.\ Aharony, Y.\ Gefen,
O.\ Entin-Wohlman, and J.\ von Delft for discussions.
This research was supported by
the German-Israeli project cooperation
(DIP --- V.K., T.H., A.W.),
European Social Fund (V.K.),
Deutsche Forschungsgemeinschaft
(FOR 723 --- C.K., V.M.; SFB 631, SGB-TR12,
De-730/3-2 --- T.H., A.W.), Nanosystems
Initiative Munich (NIM --- T.H., A.W.),
and the Israel Science Foundation (A.S.).

\newcommand{\etal}{\textit{et al.}}


\begin{thebibliography}{99}
\bibitem{Craig-etal-04}
         N.\ J.\ Craig \etal, 
         Science {\bf 304}, 565 (2004).
\bibitem{Petta-etal-05}
         J.\ R.\ Petta \etal, 
         Science {\bf 309}, 2180 (2005).
\bibitem{PJMHG04}
         J.\ R.\ Petta \etal,
         Phys.\ Rev.\ Lett.\ {\bf 93}, 186802 (2004).
\bibitem{LDiV98}
         D.Loss and D.P.DiVincenzo,
         Phys.Rev.A {\bf 57}, 120 (1998).
\bibitem{EL05}
         H.-A.\ Engel and D.\ Loss,
         Science {\bf 309}, 586 (2005).
\bibitem{SI00}
         P.\ G.\ Silvestrov and Y.\ Imry,
         Phys.\ Rev.\ Lett.\ {\bf 85}, 2565 (2000).
\bibitem{KG05}
         J.K\"onig and Y.Gefen,
         Phys.Rev.B {\bf 71}, 201308(R) (2005).
\bibitem{SSOvD05}
         M.\ Sindel \etal, 
         Phys.\ Rev.\ B {\bf 72}, 125316 (2005).
\bibitem{Special-NJP-issue}
         See, e.g., A.\ Aharony and S.\ Katsumoto (eds.),
         {\em Focus on Interference in Mesoscopic
         Systems}, New J. Phys. {\bf 9}, 111-125 (2007).
\bibitem{Ensslin02}
         S.\ Lindemann \etal,
         Phys.\ Rev.\ B {\bf 66}, 161312(R) (2002).
\bibitem{Kobayashi04}
         H.\ Aikawa \etal, 
         J.\ Phys.\ Soc.\ Jpn. {\bf 73}, 3235 (2004).
\bibitem{JMHG04}
         A.\ C.\ Johnson \etal,
         Phys.\ Rev.\ Lett. {\bf 93}, 106803 (2004).
\bibitem{BvOG05}
         R.\ Berkovits, F.\ von Oppen, and Y.\ Gefen,
         Phys.\ Rev.\ Lett. {\bf 94}, 076802 (2005).
\bibitem{Phase-lapses}
         M.\ Avinun-Kalish \etal, 
         Nature (London) {\bf 436}, 529 (2005),
         and references therein.
\bibitem{MM06}
         V.\ Meden and F.\ Marquardt,
         Phys.\ Rev.\ Lett.\ {\bf 96}, 146801 (2006).
\bibitem{LK07}
         H.-W.Lee and S.Kim,
         Phys.Rev.Lett. {\bf 98}, 186805 (2007).
\bibitem{KSAE07}
         V.\ Kashcheyevs \etal, 
         Phys.\ Rev.\ B {\bf 75}, 115313 (2007).
\bibitem{SI07}
         P.G.\ Silvestrov and Y.\ Imry,
         Phys.\ Rev.\ B {\bf 75}, 115335 (2007).
\bibitem{AYH70}
         P.\ W.\ Anderson, G.\ Yuval, and D.\ R.\ Hamann,
         Phys.\ Rev.\ B {\bf 1}, 4464 (1970).
\bibitem{NRG}
         For a recent review, see
         R.\ Bulla, T.\ Costi, and T.\ Pruschke,
         Rev.\ Mod.\ Phys.\ {\bf 80}, 395 (2008).
\bibitem{FRG}
         C.\ Karrasch, T.\ Enss, and V.\ Meden,
         Phys.\ Rev.\ B {\bf 73}, 235337 (2006).
\bibitem{LSA03}
         The derivation is analogous to that in
         Secs.~III and IV of
         E.\ Lebanon, A.\ Schiller, and F.\ B.\ Anders,
         Phys.\ Rev.\ B {\bf 68}, 155301 (2003).
\bibitem{Haldane81}
         F.\ D.\ M.\ Haldane,
         J.\ Phys. C {\bf 14}, 2585 (1981).
\bibitem{Schlottmann80II}
         See, e.g., P.\ Schlottmann, Phys.\ Rev.\ B {\bf 22},
         622 (1980) which employs linearized phase shifts.
\bibitem{Comment-on-coefficients}
         Note that
         $A = D^{1-\alpha}_{+}\bar{A}(\delta_U)$
         and
         $B = D^{1-2\beta}_{+}\bar{B}(\delta_U)$
         depend on both $\Gamma_{+}$ and $\delta_U$.
\end{thebibliography}
\end{document}